\documentclass{llncs}
\usepackage{microtype}
\usepackage{booktabs} 
\usepackage{tikz}
\usetikzlibrary{arrows,shapes,positioning}
\usepackage[linesnumbered, vlined]{algorithm2e}
\usepackage[caption=false]{subfig}
\usepackage{amsmath}
\usepackage{url}

\usepackage{xcolor}

\newcommand{\valid}{\mathit{valid}}
\newcommand{\True}{\mathit{True}}
\newcommand{\False}{\mathit{False}}
\newcommand{\Bool}{\mathit{Bool}}

\SetEndCharOfAlgoLine{}

\begin{document}
\title{Finding Regressions in Projects under Version Control Systems}

\author{Jaroslav Bend\'ik \and Nikola Bene\v s \and Ivana \v Cern\'a}

\institute{Faculty of Informatics, Masaryk University, Brno, Czech Republic\\
\email{\{xbendik,xbenes3,cerna\}@fi.muni.cz}}

\maketitle

\begin{abstract}
Version Control Systems (VCS) are frequently used to support development of
large-scale software projects. A~typical VCS repository of a~large project can
contain various intertwined branches consisting of a~large number of commits.
If some kind of unwanted behaviour (e.g.~a bug in the code) is found in the
project, it is desirable to find the commit that introduced it. Such commit is
called a~regression point. There are two main issues regarding the regression
points. First, detecting whether the project after a certain commit is
correct can be very expensive as it may include large-scale testing and/or some
other forms of verification. It is thus desirable to minimise the number of
such queries.
Second, there can be several regression points preceding the actual commit;
perhaps a bug was introduced in a certain commit, inadvertently fixed several
commits later, and then reintroduced in a yet later commit. In order to fix the
actual commit it is usually desirable to find the latest regression point.

The currently used distributed VCS contain methods for regression
identification, see e.g.~the \emph{git bisect} tool. In this paper, we present
a~new regression identification algorithm that outperforms the current tools by
decreasing the number of validity queries. At the same time, our algorithm
tends to find the latest regression points which is a feature that is missing
in the state-of-the-art algorithms. The paper provides an experimental
evaluation of the proposed algorithm and compares it to the state-of-the-art
tool \emph{git bisect} on a real data set.
\end{abstract}

\section{Introduction}

\emph{Version Control Systems} (VCS) have become ubiquitous in the area of (not
only) software development, from small toy projects to large-scale industrial
ones.  The recent years saw a rise in the popularity of \emph{Distributed VCS}
such as \emph{git}~\cite{git}, \emph{bazaar}~\cite{bazaar}, \emph{Mercurial}~\cite{mercurial} and
many others. These allow for almost seamless cooperation of a~large number of
developers and support extensive project branching and merging of branches.
After a~project has been in the development process for some time, the commit
graph of its repository may grow to be very large.

As projects grow larger, the appearance of bugs (i.e.~unwanted behaviour of the
developed product) is going to be inevitable. Software bugs can be usually
caught early if the development teams employ extensive testing techniques
(unit tests, performance regression tests, etc.); however, from time to time
a~bug, or a commit that changed properties of the project, may
creep into the VCS repository and lie there undetected for some time.
Such bug is usually discovered by e.g.~extending the coverage of the tests or
by employing some other verification technique such as model
checking~\cite{bible}. In order to fix the bug it is very useful to identify
the commit that introduced  the bug as this  commit  typically contains a
relatively small set of source code changes. It is much easier to properly
understand and fix a bug when you only need to check a very small set of
changes of the source code. Sometimes we are not looking for the commit that
introduced a bug, but rather for a~commit that caused a change between
some ``old'' and ``new'' state of the project.
As an example, we might be looking for the commit that introduced a~particular
fix.  In such cases it can seem
confusing to use the terms ``correct'' and ``buggy'' to refer to the state before
and after the change, respectively. We thus instead use the terms
\emph{valid} and \emph{invalid} commit; we further use the term
\emph{regression point} to denote the point where the property of interest has
been changed.

The problem of finding regression points has been addressed before and there
have been developed tools for solving this problem, such as \emph{git
bisect}~\cite{git_bisect_doc}. These tools have proved themselves to be very
useful and are commonly used during software development nowadays. Yet, there
are several issues related to finding regression points and only some of them
are targeted by the state-of-the-art tools.

First, the search for regression points consists of several queries of the
form: ``Given a~certain commit, is the bug present in the system after this
commit?'' Such queries, which we call \emph{validity queries}, may consist of
several expensive tasks like running tests, model checking, code inspection, or
other forms of verification. It~is thus desirable to minimise the number of
these queries. Second, the validity of commits does not, in general, have to be
monotone. Perhaps a bug was introduced in a~certain commit, inadvertently fixed
several commits later, and then reintroduced in a~yet later commit.
This means that there are possibly several regression points preceding the
actual invalid commit, but only the identification of the latest regression
point can usually help us to fix the bug.
The third issue concerns large projects with many branches.
If,~for example, a new test case is employed, then more than just one active
branch can fail the test case and it is desirable to identify a regression
point for each of these branches.  We can find the regression point for each
branch separately. However, dealing with all branches simultaneously can save
some validity queries and thus optimise the search.

The state-of-the-art tools target the need to minimise the number of validity
queries that are performed. However, they do not tend towards identification of
latest regression points, and they deal with only  a single invalid branch at
a~time.

The goal of this paper is to provide a~novel algorithm, called the Regression
Predecessors Algorithm (RPA), that solves the problem of finding regression
points in VCS repositories. The algorithm minimises the number of validity
queries and at the same time tries to find the latest regression points both for
single and multiple invalid branches.  RPA has several variants which we
compare on a~set of real open-source projects. Moreover, we compare the RPA
algorithm with the state-of-the-art tool git bisect and demonstrate that
our algorithm outperforms git bisect in all of the three above-mentioned
criteria: in the number of performed validity queries, in finding the latest
regression points, and in finding regression points for multiple invalid
branches.

\medskip
The rest of the paper is organised as follows. Section~\ref{sec:prelim} defines
basic notions and states the problem formally. Section~\ref{sec:algo} presents
the Regression Predecessors Algorithm and illustrates its behaviour on a small
example. Section~\ref{sec:related} reviews the related work and compares RPA
with other known algorithms. Section~\ref{sec:exp} gives an experimental
evaluation of different variants of RPA and compares RPA with the
state-of-the-art tool git bisect on a set of real benchmarks.

\section{Preliminaries and Problem Formulation}\label{sec:prelim}

\begin{definition}
A \emph{rooted directed acyclic graph} is a directed graph  $G = (V,E)$ with
exactly one root (i.e.~a~vertex with no incoming edges) and with no cycle
(i.e.~there is no path $\langle v_0,v_1, \cdots v_k \rangle$ in the graph such
that $v_0 = v_k$ and $k > 0$).
A \emph{rooted annotated directed acyclic graph (RADAG)} is a pair $(G,
\valid)$, where $G$ is a~rooted directed acyclic graph with root $r$ and
$\valid: V \rightarrow \Bool$ is a~\emph{validation function} satisfying
$\valid(r) = \True$.

\end{definition}

We use RADAGs to model structures which arise from using version control
systems (VCS). Each vertex in $G$ corresponds to a~commit in VCS repository.
An edge between two vertices represents two subsequent commits. The root
corresponds to the initial commit and the leaves (vertices with no outgoing
edges) correspond to the latest commits of individual branches.

The validation function expresses whether a particular commit has the desired
property that the system after this commit is correct (i.e.~does not contain
the bug). We call the vertices with this property (for which the validation function evaluates to $\True$) \emph{valid} vertices and the others  \emph{invalid} ones.
Note that we assume that the graph has only one root and that the root is valid.
If this is not the case, the graph can be easily modified  by adding a dummy initial valid commit. 

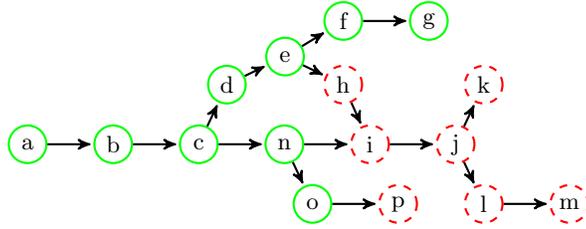
\begin{figure}[t]
	\centering
%

\begin{tikzpicture}[->,>=stealth',shorten >=1pt,auto,node distance=1cm,minimum size=0.5cm,inner sep=0pt,
  thick,every node/.style={circle,draw},
  valid/.style={circle,draw,draw=green},  
  invalid/.style={circle,dashed,draw,draw=red},
  v/.style={fill=green},
  i/.style={fill=red}]

\node[valid] (a) {a};
\node[valid] (b) [right = 0.61 of a] {b};
\node[valid] (c) [right = 0.61 of b] {c};
\node[valid] (n) [right = 0.61 of c] {n};
\node[invalid] (i) [right = 0.61 of n] {i};
\node[valid] (d) [above right = 0.42 and -0 of c] {d};
\node[valid] (e) [above right = 0 and 0.4 of d] {e};
\node[invalid] (h) [below right = 0 and 0.4 of e] {h};

\node[valid] (o) [below right = 0.42 and -0 of n] {o};
\node[invalid] (p) [right = 0.61 of o] {p};

\node[invalid] (j) [right = 0.61 of i] {j};
\node[invalid] (k) [above right = 0.42 and -0 of j] {k};
\node[invalid] (l) [below right = 0.42 and -0 of j] {l};
\node[invalid] (m) [right = 0.61 of l] {m};

\node[valid] (f) [above right = 0.1 and 0.4 of e] {f};
\node[valid] (g) [right = 0.61 of f] {g};

\path[every node/.style={font=\sffamily\small}]
(a) 	edge node [left] {} (b)
(b) 	edge node [left] {} (c)
(c) 	edge node [left] {} (d)
		edge node [left] {} (n)	
(d) 	edge node [left] {} (e)
(e) 	edge node [left] {} (f)
		edge node [left] {} (h)
(f) 	edge node [left] {} (g)
(h) 	edge node [left] {} (i)
(i) 	edge node [left] {} (j)
(j) 	edge node [left] {} (k)
	 	edge node [left] {} (l)
(l) 	edge node [left] {} (m)
(n) 	edge node [left] {} (o)
		edge node [left] {} (i)
(o) 	edge node [left] {} (p)
;
\end{tikzpicture}
	\caption{An example of a RADAG, the dashed vertices are invalid. There are three invalid leaves in this example: $k$, $m$, $p$, and four regression points: $(c,d), (e,h), (n,i)$ and $(o,p)$. The first three   regression points are regression predecessors of invalid leaves $k, m$; the fourth regression point is the regression predecessor of the invalid leaf $p$.}
  	\label{fig:radag}
\end{figure}

\begin{definition}
A~\emph{regression point} of a~RADAG $((V,E), \valid)$ is a~pair of vertices
$(u,v)$ such that $(u,v) \in E$, $\valid(u) = \True$, and $\valid(v) = \False$.
A regression point $(u,v)$ is a \emph{regression predecessor} of a vertex $w
\in V$ if $w$ is equal to $v$ or $w$ is reachable from $v$.
\end{definition}

We are now ready to formally state our problem.

\textbf{Regression Predecessors Problem:}
{Given} a~RADAG $(G, \valid)$ and a set of invalid leaves $L$ of $G$, {find} at
least one regression predecessor for each leaf from~$L$.

Note that one invalid leaf can have several regression predecessors and one regression point can be a regression predecessor of several invalid leaves. Therefore, the regression predecessors  problem may have several different solutions.

\section{Regression Predecessors Algorithm (RPA)}\label{sec:algo}

A naive solution to the regression predecessors problem would be to   evaluate
the function $\valid$ for each vertex, identify all regression points in
$G$, and find a regression predecessor of each invalid leaf  using the
reachability relation.
Because this approach identifies all regression points we can choose the latest
regression predecessor of every invalid leaf. However, the price is crucial;
the function $\valid$ is evaluated for every vertex which is assumed to be extremely time-consuming.

In this section we present a new algorithm, the Regression
Predecessors Algorithm (RPA), that substantially decreases the
number of vertices for which the function $\valid$ is evaluated and
tends to find the latest regression points at the same time.

\subsection{Basic Schema}

The main idea of RPA is based on the observation that if a leaf $l$ is invalid then every path starting in a valid vertex and leading to $l$ must  contain at least one regression predecessor of $l$. This reduces the problem to two tasks: finding a path and detecting a regression point on the path.

\begin{algorithm}[t]
\SetKwInput{Input}{input}\SetKwInput{Output}{output}
\SetKwFunction{buildPath}{buildPath}
\SetKwFunction{getRegressionPoint}{getRegressionPoint}
\SetKwFunction{propagateRegressionPoint}{propagateRegressionPoint}
\SetKwFunction{getUnprocessedLeaf}{getUnprocessedLeaf}

\SetKwProg{Fn}{function}{}{}\SetKwFunction{rpa}{RPA}%
\Fn(){\rpa{G, L}}{

\Input{a RADAG $G = ((V,E), \valid: V \rightarrow \Bool)$ with root $r$}
\Input{a set of invalid leaves $L$}
\Output{a regression predecessor for each leaf $l \in L$}
\BlankLine

	$\mathit{UnprocessedLeaves} \gets L$\;
	$\mathit{KnownValid} \gets \{r \}$\;
	
	\While{$UnprocessedLeaves \neq \emptyset$}{
		$l \gets$ \emph{a leaf from $\mathit{UnprocessedLeaves}$}\;
		$\mathit{UnprocessedLeaves} \gets \mathit{UnprocessedLeaves} \setminus \{l\}$\;
		$p_l \gets$ \emph{a path $\langle x, \ldots, l \rangle$ such that $x \in \mathit{KnownValid}$} \;
		$(u,v) \gets$ \emph{find a regression point on $p_l$}\;
		$\mathit{KnownValid} \gets \mathit{KnownValid} \cup \mathit{NewValid}$ \; 
		\textbf{output} \emph{$(u,v)$ is a regression predecessor of $l$}\;
		$\propagateRegressionPoint(v, (u,v))$ \tcp*{optional, Alg.~2}
	}
}
 \caption{Regression Predecessors Algorithm (basic schema)}
\label{alg-rpa}
\end{algorithm}

For the basic description of RPA see Algorithm~\ref{alg-rpa}.
The algorithm maintains the set  $\mathit{UnprocessedLeaves}$ which consists of those
invalid leaves for which a regression predecessor has not been computed yet.
The set $\mathit{KnownValid}$ consists of those vertices for which the function
$\valid$ has been evaluated and are valid (initially only the \emph{root} is
known to be valid). In each iteration, the algorithm chooses a   leaf  $l$ from
the set $\mathit{UnprocessedLeaves}$. A regression predecessor of $l$ is acquired by
building a path $p_l$ which connects a~valid vertex $x \in \mathit{KnownValid}$ with
$l$ and by finding a regression point $(u,v)$ on this path. While searching for
the regression point $(u,v)$, the function $\valid$ is evaluated  for some
vertices on the path~$p_l$. The newly detected valid vertices form a set
$\mathit{NewValid}$ and the set $\mathit{KnownValid}$ is updated accordingly.

The algorithm also exploits the fact that one regression point can be a
regression predecessor of several invalid leaves. Therefore, every time a
regression point $(u,v)$ is found, it is  {propagated} to every invalid leaf $m
\in \mathit{UnprocessedLeaves}$ such that $m$ reachable from $v$. Every such $m$ is
removed from $\mathit{UnprocessedLeaves}$ (see  Algorithm~\ref{alg-propagation}).
After this propagation step, the procedure also removes from the graph all vertices
reachable from $v$.   By  {removing} vertices we avoid propagation of
regression points to leaves for which  a regression predecessor   has already
been found and avoid unnecessary traversal of the graph.

On the one hand, the propagation can result in saving some validation calls.
On the other hand, the use of propagation may be in conflict with the desire to
identify the latest regression predecessors.
Therefore, the usage of propagation is optional.
Section~\ref{sec:exp} demonstrates the behaviour of the algorithm both with and
without the propagation step.

There are further three key aspects that affect the efficiency
of the algorithm: the order in which leaves are chosen from the set
$\mathit{UnprocessedLeaves}$, the method of building the path connecting a valid vertex
with the invalid leaf, and the method of regression points identification.
We focus on these three aspects in the following text.

\begin{algorithm}[t]
\SetKwInput{Input}{input}\SetKwInput{Output}{output}
\SetKwProg{Fn}{function}{}{}\SetKwFunction{prop}{propagateRegressionPoint}%
\Fn(){\prop{k, (u,v)}}{
\Input{a regression point $(u,v)$}
\Input{a vertex $k$ reachable from $v$ (or $v = k$)}
\BlankLine
	\For{$(k,l) \in E$}{
		$\prop{l, (u,v)}$\;
	}
	\If{$k \in \mathit{UnprocessedLeaves}$}{
		\textbf{output} \emph{$(u,v)$ is a regression predecessor of $k$}\;
		$\mathit{UnprocessedLeaves} \gets \mathit{UnprocessedLeaves} \setminus \{k\}$
	}
	\emph{remove $k$ from the graph}
}	

\caption{regression point propagation}
\label{alg-propagation}
\end{algorithm}

\subsection{Identification of Regression Points}
In this subsection we give the details of our solution to the problem of
finding a regression point on a given path $p = \langle v_0, v_1,
\ldots, v_l \rangle$  connecting a valid vertex $v_0$ with an invalid vertex
$v_l$.

\paragraph{Linear search}
The simplest solution to the task is to evaluate the function $\valid$ for each
vertex on the path, starting with $v_l$ and going backwards. As soon as a~valid
vertex $v_i$ is found, the algorithm outputs $(v_{i}, v_{i+1})$ as a~regression
predecessor of $v_l$. By~starting with $v_l$ and going backwards we guarantee
that $(v_{i}, v_{i+1})$ is the nearest regression point of $v_l$ along this
path. The disadvantage of this approach is that in the worst case all vertices
on the path are tested for validity. Because the commit graphs of VCS usually
contain hundreds or thousands of commits and the evaluation of the function
valid is assumed to be very expensive, the linear search is practically
unusable.

\paragraph{Binary search}

Provided that the first vertex of the path is valid and the last is
invalid (which is always our case) we can use binary search to find a regression point.
Let $p = \langle v_0, v_1, \ldots v_{mid}, \ldots, v_l \rangle$ be a path such
that $v_0$ is valid, $v_l$ is invalid, and $v_{mid}$ is the middle vertex of
this path. If $v_{mid}$ is valid then there is a regression point on the path
$\langle v_{mid}, \ldots, v_l \rangle$. Otherwise, there is a regression point
on the path $\langle v_0, \ldots v_{mid} \rangle$. We can thus always reduce
$p_l$ into half and recursively repeat the procedure.

Contrary to the linear search approach it is not guaranteed that the binary
search finds a regression point which is nearest to $v_l$. The main advantage
of the binary search is that it evaluates the function $\valid$ only for
logarithmically many vertices on the path~$p$, which is optimal.

\paragraph{Multiplying search}
The so-called multiplying search approach combines the advantages of both binary and linear search
approaches as it performs asymptotically fewer validity checks than the linear
search and at the same time tends to find a regression point which is closer
to the last vertex $v_l$ than the regression point found by the binary search.

Let $p = \langle v_0, v_1, \ldots, v_l \rangle$ be a path such that $v_0$ is valid and $v_l$ is invalid. The \emph{multiplying search} first evaluates the function $\valid$ for  the vertex $v_{l-1}$. If  $v_{l-1}$ is not valid, then the function is stepwise evaluated for vertices    $v_{l-2^0-2^1}$, $v_{l-2^0-2^1-2^2}$, $v_{l-(2^4-1)}, \ldots$     forming exponentially large gaps between individual vertices.  The procedure eventually finds an~$i$  such that  the vertex $v_{l - (2^{i-1}-1)}$ is invalid and either  $v_{l-(2^i-1)}$ is valid or $l-(2^i-1) <0$. If~the former happens, then  the procedure  recursively continues with the new path $\overline{p} = \langle v_{l-(2^i-1) }, v_{l-(2^i-2)}, \ldots,$ $ v_{l - (2^{i-1}-1)} \rangle$.
In the latter case the procedure recursively continues with the path $\overline{p} = \langle v_0, v_1, \ldots, v_{l -(2^{i-1}-1)} \rangle$. The procedure converges to a path containing only two vertices such that the first vertex of the path is valid and the second invalid, i.e., a regression point is found. For the complete description   see Algorithm~\ref{alg-multiplying-search}.

The number $C(n)$ of vertices on which the function $\valid$ is evaluated on
a path of length $n$ is bounded by the recurrence equation $C(n) \leq
C(\frac{n}{2}) + \log n$. In each recursive call  the number of evaluations
is at most $\log{n}$ and  the length of the path  is  decreased  at least by
half. The solution of the recurrence equation (using the Master
theorem~\cite{master_theorem}) gives an upper bound $\mathcal{O}(\log^2{n})$ on
the number of vertices on which the function $\valid$ is evaluated.

\begin{algorithm}[t]
\SetKwInput{Input}{input}\SetKwInput{Output}{output}
\SetKwFunction{findRegressionPoint}{findRegressionPoint}
\SetKwFunction{propagateRegressionPoint}{propagateRegressionPoint}
\SetKwFunction{updatePaths}{updatePaths}

\SetKwProg{Fn}{function}{}{}\SetKwFunction{mult}{multSearch}%
\Fn(){\mult{p}}{
\Input{a path $p =\langle v_0, \ldots, v_l \rangle$ with valid $v_0$ and invalid $v_l$}
\Output{a regression point contained in $p$}
\BlankLine

	\If{$l = 1$}{
		\Return $(v_0, v_1)$ 
	}
	$k = 1$\;
	\While{$l-(2^k - 1) > 0$}{ 
		\If{$\valid(v_{l-(2^k - 1)})$}{
			\Return $\mult(\langle v_{l-(2^k - 1)}, \ldots, v_{l-(2^{k-1} - 1)} \rangle)$
		}
		$k = k + 1$\;
	}
	\Return $\mult(\langle v_0, \ldots, l-(2^{k-1} - 1) \rangle)$\;
}
 \caption{multiplying search}
\label{alg-multiplying-search}
\end{algorithm}

\subsection{Leaf Selection and Path Construction}
Our next goal is to specify the order in which unprocessed leaves are chosen
and determine the method of building a path connecting a valid vertex with the
chosen leaf.

We assume that the directed acyclic graphs induced by VCS are represented using
adjacency lists (see~\cite{cormen}) in which every vertex is equipped both
with a~list of its direct successors and a~list of its direct predecessors.
In the initialisation phase of RPA we compute the length of the shortest
paths from $v$ to $l$ for each vertex $v \in V$ and invalid leaf $l \in L$. For
every pair $(v, l) \in V \times L$ we maintain a successor of $v$ so that the
chain of successors originating at the vertex $v$ runs forward along a shortest
path from $v$ to $l$. This computation is done by running a~backwards
breadth-first search from each $l \in L$ using the list of predecessors, for
details see e.g.~\cite{cormen}.

In what follows we use $dist(v,l)$ to denote the length of the shortest path
leading from the vertex $v$ to the leaf $l$; we further define:
\begin{align*}
dist(l) &= min\{ dist(u,l) \mid u \in \mathit{KnownValid}\}\\
start(l) &= u \text{ such that } u \in \mathit{KnownValid}\\
		&\, \, \, \, \, \, \text{ and } dist(u,l) = dist(l).
\end{align*}
In other words, $dist(l)$ denotes the length of a shortest path leading to $l$
from a vertex $u$ for which the function $\valid$ has been evaluated and is
valid (i.e.~belongs to the set $\mathit{KnownValid}$). The first vertex of such a path
is denoted $start(l)$.
As the set $\mathit{KnownValid}$ changes during the computation, so may the values
$dist(l)$ and $start(l)$. Initially, only the root $r$ of the graph is known to
be valid, therefore $dist(l) = dist(r,l)$ and $start(l) = r$ for each $l \in L$.

\begin{algorithm}[t]

\SetKwInput{Input}{input}\SetKwInput{Output}{output}
\SetKwFunction{buildPath}{buildShortestPath}
\SetKwFunction{getRegressionPoint}{getRegressionPoint}
\SetKwFunction{propagateRegressionPoint}{propagateRegressionPoint}
\SetKwFunction{getUnprocessedLeaf}{getUnprocessedLeaf}
\SetKwFunction{computeSP}{computeShortestPaths}
\SetKwFunction{updatePriority}{updatePriorities}

\SetKwProg{Fn}{function}{}{}\SetKwFunction{rpa}{priorityBasedRPA}%
\Fn(){\rpa{G, L}}{
\Input{a RADAG $G = ((V,E), \valid: V \rightarrow \Bool)$ with root $r$}
\Input{a set of invalid leaves $L$}
\Output{a regression predecessor for each leaf $l \in L$}
\BlankLine
	\emph{for each $(v,l) \in V \times L$ compute the value  $dist(v,l)$ }\;
	\For{ \emph{each} $l \in L$ }{
		$dist(l) \gets dist(r,l)$\;
		$start(l) \gets r$\;
	}	
		
	$\mathit{UnprocessedLeaves} \gets L$ \tcp*{priority queue}
	
	\While{$\mathit{UnprocessedLeaves} \neq \emptyset$}{
		$l \gets \mathit{UnprocessedLeaves.dequeueMinimum()}$\;
		$p_l \gets$ \emph{a shortest path $\langle x, \ldots, l \rangle$ such that $x = \mathit{start(l)}$} \;
		$(u,v) \gets$ \emph{find a regression point on $p_l$}\;
		$\mathit{KnownValid} \gets \mathit{KnownValid} \cup \mathit{NewValid}$\;
		\textbf{output} \emph{$(u,v)$ is the regression predecessor of $l$}\;
		$\propagateRegressionPoint(v, (u,v))$ \tcp*{optional, Alg.2}
		$\updatePriority(\mathit{NewValid})$ \tcp*{Alg.5}
	}
}	
 \caption{Regression Predecessor Algorithm}
\label{alg-priority-rpa}
\end{algorithm}

The way in which RPA fixes the order in which invalid leaves are processed and determines which paths should be used for identification of regression points is based on the following observation. The shorter path
we process the fewer number of evaluations of the function $\valid$ is
performed, independent on the regression finding approach.
For a complete description of the RPA algorithm see Algorithm~\ref{alg-priority-rpa},
for an illustrative example see Sec.~\ref{example}.

The RPA algorithm maintains the set $\mathit{UnprocessedLeaves}$ as a priority
queue where each $l \in \mathit{UnprocessedLeaves}$ is assigned the priority $dist(l)$.
In every iteration the algorithm extracts the leaf $l$ with minimum priority
from $\mathit{UnprocessedLeaves}$ and constructs the shortest path leading to $l$.
Moreover, each iteration is supplemented by the method
$\mathit{updatePriorities(newValid)}$ that updates the $dist(l)$ and $start(l)$ values
(see Algorithm~\ref{alg-update-priorities}).

\begin{algorithm}[t]
 \SetKwInput{Input}{input}\SetKwInput{Output}{output}

\SetKwProg{Fn}{function}{}{}\SetKwFunction{up}{updatePriorities}%
\Fn(){\up{NewValid}}{

\Input{A set of valid vertices $NewValid$}
\BlankLine
	\For{\emph{each} $v$ \emph{in} $NewValid$}{
		\For{\emph{each} $leaf \in UnprocessedLeaves$}{
			\If{$dist(v,leaf) < dist(leaf)$}{
				$start(leaf) \gets v$\;
				$dist(leaf) \gets dist(v,leaf)$
			}
		}
	}
}	

 \caption{priority update}
\label{alg-update-priorities}
\end{algorithm}

\subsection{Example} \label{example}
%

\tikzstyle{op}=[opacity=.2]
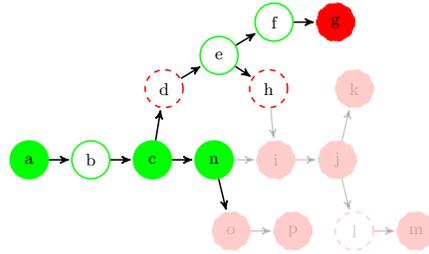
\begin{figure*}[!ht]
\noindent
\begin{minipage}{0.5\textwidth}
\textbf{I. iteration}\\
-- Removed vertices = $\emptyset$\\
-- $KnownValid$ = $\{a\}$\\
-- Priority queue = $\langle p,g,k,m \rangle$\\
-- $dist(p) = 5$, $dist(g) = 6$,\\ $dist(k) = 6$, $dist(m) = 7$\\
-- $start(p) = a$\\
-- path $p_p$ = $\langle a,b,c,n,o,p \rangle$\\
-- Tested vertices: $o,c,n$\\
-- Regression point of $p_p$ = $(n,o)$\\
-- Propagated to: $\{p\}$\\
\end{minipage}
\begin{minipage}{0.4\textwidth}
	\scalebox{0.72}{
%

\begin{tikzpicture}[->,>=stealth',shorten >=1pt,auto,node distance=1cm,minimum size=0.7cm,
  thick,every node/.style={circle,draw},
  valid/.style={circle,draw,draw=green},  
  invalid/.style={circle,dashed,draw,draw=red},
  v/.style={fill=green},
  i/.style={fill=red}]

\node[valid,v] (a) {a};
\node[valid] (b) [right = 0.41 of a] {b};
\node[valid] (c) [right = 0.41 of b] {c};
\node[valid] (n) [right = 0.41 of c] {n};
\node[invalid] (i) [right = 0.41 of n] {i};
\node[invalid] (d) [above right = 0.8 and -0.3 of c] {d};
\node[valid] (e) [above right = 0.1 and 0.5 of d] {e};
\node[invalid] (h) [below right = 0.1 and 0.4 of e] {h};

\node[invalid] (o) [below right = 0.8 and -0.2 of n] {o};
\node[invalid,i] (p) [right = 0.41 of o] {p};

\node[invalid] (j) [right = 0.41 of i] {j};
\node[invalid,i] (k) [above right = 0.8 and -0.2 of j] {k};
\node[invalid] (l) [below right = 0.8 and -0.2 of j] {l};
\node[invalid,i] (m) [right = 0.41 of l] {m};

\node[valid] (f) [above right = 0.1 and 0.5 of e] {f};
\node[invalid,i] (g) [right = 0.41 of f] {g};

\path[every node/.style={font=\sffamily\small}]
(a) 	edge node [left] {} (b)
(b) 	edge node [left] {} (c)
(c) 	edge node [left] {} (d)
		edge node [left] {} (n)	
(d) 	edge node [left] {} (e)
(e) 	edge node [left] {} (f)
		edge node [left] {} (h)
(f) 	edge node [left] {} (g)
(h) 	edge node [left] {} (i)
(i) 	edge node [left] {} (j)
(j) 	edge node [left] {} (k)
	 	edge node [left] {} (l)
(l) 	edge node [left] {} (m)
(n) 	edge node [left] {} (o)
		edge node [left] {} (i)
(o) 	edge node [left] {} (p)
;
\end{tikzpicture}
	}
\end{minipage}

\vspace{0pt}
\noindent
\begin{minipage}{0.5\textwidth}
\textbf{II. iteration}\\
-- Removed vertices = $\{o,p\}$\\
-- $KnownValid$ = $\{a,c,n\}$\\
-- Priority queue = $\langle k,m,g \rangle$\\
-- $dist(k) = 3$, $dist(m) = 4$, $dist(g) = 4$\\
-- $start(k) = n$\\
-- path $p_k$ = $\langle n,i,j,k \rangle$\\
-- Tested vertices: $j,i$\\
-- Regression point of $p_k$ = $(n,i)$\\
-- Propagated to: $\{k,m\}$\\
\end{minipage}
\begin{minipage}{0.4\textwidth}
	\scalebox{0.72}{
%

\begin{tikzpicture}[->,>=stealth',shorten >=1pt,auto,node distance=1cm,minimum size=0.7cm,
  thick,every node/.style={circle,draw},
  valid/.style={circle,draw,draw=green},  
  invalid/.style={circle,dashed,draw,draw=red},
  v/.style={fill=green},
  i/.style={fill=red},
  c/.style={}]

\node[valid,v] (a) {a};
\node[valid] (b) [right = 0.41 of a] {b};
\node[valid,v] (c) [right = 0.41 of b] {c};
\node[valid,v] (n) [right = 0.41 of c] {n};
\node[invalid] (i) [right = 0.41 of n] {i};
\node[invalid] (d) [above right = 0.8 and -0.3 of c] {d};
\node[valid] (e) [above right = 0.1 and 0.5 of d] {e};
\node[invalid] (h) [below right = 0.1 and 0.4 of e] {h};

\node[invalid,i,op] (o) [below right = 0.8 and -0.2 of n] {o};
\node[invalid,i,op] (p) [right = 0.41 of o] {p};

\node[invalid] (j) [right = 0.41 of i] {j};
\node[invalid,i] (k) [above right = 0.8 and -0.2 of j] {k};
\node[invalid] (l) [below right = 0.8 and -0.2 of j] {l};
\node[invalid,i] (m) [right = 0.41 of l] {m};

\node[valid] (f) [above right = 0.1 and 0.5 of e] {f};
\node[invalid,i] (g) [right = 0.41 of f] {g};

\path[every node/.style={font=\sffamily\small}]
(a) 	edge node [left] {} (b)
(b) 	edge node [left] {} (c)
(c) 	edge node [left] {} (d)
		edge node [left] {} (n)	
(d) 	edge node [left] {} (e)
(e) 	edge node [left] {} (f)
		edge node [left] {} (h)
(f) 	edge node [left] {} (g)
(h) 	edge node [left] {} (i)
(i) 	edge node [left] {} (j)
(j) 	edge node [left] {} (k)
	 	edge node [left] {} (l)
(l) 	edge node [left] {} (m)
(n) 	edge node [left] {} (o)
		edge node [left] {} (i)
(o) 	edge[op] node [left] {} (p)
;
\end{tikzpicture}
	}
\end{minipage}

\noindent
\begin{minipage}{0.5\textwidth}
\textbf{III. iteration}\\
-- Removed vertices = $\{o,p,i,j,k,l,m\}$\\
-- $KnownValid$ = $\{a,c,n\}$\\
-- Priority queue = $\langle g \rangle$\\
-- $dist(g) = 4$\\
-- $start(g) = c$\\
-- path $p_g$ = $\langle c,d,e,f,g \rangle$\\
-- Tested vertices: $f$\\
-- Regression point of $p_g$ = $(f,g)$\\
-- Propagated to: $\{g\}$\\
\end{minipage}
\begin{minipage}{0.4\textwidth}
	\scalebox{0.72}{
%

\begin{tikzpicture}[->,>=stealth',shorten >=1pt,auto,node distance=1cm,minimum size=0.7cm,
  thick,every node/.style={circle,draw},
  valid/.style={circle,draw,draw=green},  
  invalid/.style={circle,dashed,draw,draw=red},
  v/.style={fill=green},
  i/.style={fill=red},
  c/.style={}]

\node[valid,v] (a) {a};
\node[valid] (b) [right = 0.41 of a] {b};
\node[valid,v] (c) [right = 0.41 of b] {c};
\node[valid,v] (n) [right = 0.41 of c] {n};
\node[invalid,op,i] (i) [right = 0.41 of n] {i};
\node[invalid] (d) [above right = 0.8 and -0.3 of c] {d};
\node[valid] (e) [above right = 0.1 and 0.5 of d] {e};
\node[invalid] (h) [below right = 0.1 and 0.4 of e] {h};

\node[invalid,i,op] (o) [below right = 0.8 and -0.2 of n] {o};
\node[invalid,i,op] (p) [right = 0.41 of o] {p};

\node[invalid,i,op] (j) [right = 0.41 of i] {j};
\node[invalid,i,op] (k) [above right = 0.8 and -0.2 of j] {k};
\node[invalid,op] (l) [below right = 0.8 and -0.2 of j] {l};
\node[invalid,i,op] (m) [right = 0.41 of l] {m};

\node[valid] (f) [above right = 0.1 and 0.5 of e] {f};
\node[invalid,i] (g) [right = 0.41 of f] {g};

\path[every node/.style={font=\sffamily\small}]
(a) 	edge node [left] {} (b)
(b) 	edge node [left] {} (c)
(c) 	edge node [left] {} (d)
		edge node [left] {} (n)	
(d) 	edge node [left] {} (e)
(e) 	edge node [left] {} (f)
		edge node [left] {} (h)
(f) 	edge node [left] {} (g)
(h) 	edge[op] node [left] {} (i)
(i) 	edge[op] node [left] {} (j)
(j) 	edge[op] node [left] {} (k)
	 	edge[op] node [left] {} (l)
(l) 	edge[op] node [left] {} (m)
(n) 	edge node [left] {} (o)
		edge[op] node [left] {} (i)
(o) 	edge[op] node [left] {} (p)
;
\end{tikzpicture}
	}
\end{minipage}

\caption{An~illustrative example}
\label{fig:exec}
\end{figure*}

Figure~\ref{fig:exec} demonstrates the execution of RPA with multiplying search and propagation.
In this example, the RADAG has only invalid leaves, $L = \{g,k,m,p\}$, and the
task is to find the regression predecessors for all leaves. The computation
consists of 3 iterations. The function $\valid$ is evaluated only for 7 out of
16 vertices and 3 regression points are found.
We list the values of control variables in each iteration and illustrate them
on the graph. The vertices on which the $\valid$ function has been evaluated
are filled with green or red color depending on their validity. The vertices
removed from the graph are shaded.

\subsection{Related Work}\label{sec:related}
To the best of our knowledge, the first tool for finding regression points was
the \emph{git bisect} tool~\cite{git_bisect_doc} which is a part of the
distributed VCS \emph{git}~\cite{git}. The method used in the git bisect tool
is called \emph{bisection} and it was subsequently adopted by other VCS like
Mercurial~\cite{mercurial}, Subversion~\cite{svn}, and Bazaar~\cite{bazaar}.

The bisection algorithm takes as an input a single invalid commit and finds
a~regression point that precedes this commit.
We give only a~brief description of the algorithm here, more elaborated
description can be found in~\cite{git_bisect_alg}.
The algorithm represents the commits using a~directed acyclic graph and assumes
that the function $\valid$ is monotone, i.e.~that every successor of an invalid
commit is also invalid.
It starts by taking as an input a single invalid commit called ``bad'' together
with a one or more commits which are known to be valid.
Then, it iteratively repeats the following steps:
\begin{itemize}
\item[(i)] Keep only the commits that: a) precede ``bad'' commit (including the
``bad'' commit itself) and b) do not precede a commit which is known to be
valid (excluding the commits which are known to be valid).
\item[(ii)] Associate to each commit $c$ a number
  $r = min\{(x + 1), n - (x + 1)\}$
where $x$ is the number of commits that precede the commit $c$ and $n$ is the
total number of commits in the graph. Roughly speaking, this number represents
the amount of information that can be obtained by evaluating the function
$\valid$ on $c$. If $c$ is valid then all of its predecessors are valid (based
on the assumption that the function $\valid$ is monotone), otherwise all of its
successors are invalid.
\item[(iii)] Evaluate the function $\valid$ for the commit $v$ with the highest
associated number. If $v$ is invalid then it becomes the ``bad'' commit.
\end{itemize}

Eventually there will be only one invalid commit left in the graph with one of
its predecessors in the original graph being valid. This pair of vertices forms
the regression predecessor of the original ``bad'' commit.  Although the main
idea of the bisection method is based on the monotonicity of the function
$\valid$, it is guaranteed that the algorithm finds a regression predecessor of
the ``bad'' commit even if the function $\valid$ is not monotone.

There are two main drawbacks of git bisect comparing to RPA.
First, the bisection algorithm does not tend to find the latest regression
predecessor.
Second, experiments (see the following section) demonstrate that git bisect
evaluates more commits than RPA. The reason of this behaviour is that RPA
prefers shortest paths while git bisect prefers vertices with the highest
associated number. To demonstrate the difference let us consider a  graph with
one leaf and two paths  connecting the root with the leaf.  If one path is very
short and the second one very long, then RPA prefers the short path while git
bisect evaluates vertices on the long one. If a graph contains only one path
leading to an invalid leaf, git bisect evaluates the same vertices as RPA
combined with binary search.

There is also further related work that deals with problems similar to ours.
Heuristics for automated \emph{culprit finding}~\cite{culprit} are used for
isolating one or more code changes which are suspected of causing a code
failure in a sequence of project versions.  They assume that the codebase is
tested/validated regularly (e.g.~after every $n$ commits) using some test suit.
If a bug is detected, they search for the culprit only among the changes to the
codebase that have been made since the latest appliance of the test suite. The
individual versions are rated according to their potential to cause the failure
(e.g.~versions with many code changes are rated higher) and   versions with
high rate are tested as first. The \emph{culprit finding}
technique~\cite{culprit} is efficiently applicable only for searching in a
short term history and it assumes that there is only one culprit.

\emph{Delta debugging}~\cite{delta_debugging} is a methodology to automate the
debugging of programs using the approach of a~hypothesis-trial-result loop. For
a~given code and a test case that detects a bug in the code, the delta
debugging algorithm can be used to trim useless functions and lines of the code
that are not needed to reproduce to bug. The delta debugging cannot be used for
finding regression points in VCS. However, we believe that it can be
incorporated into RPA and improve its performance by reducing the portion of
code that need to be validated by the function $\valid$.

A \emph{regression testing}~\cite{regression_testing} and \emph{continuous integration testing}~\cite{cit} 
are types of software testing that verifies that software previously developed and
tested still performs correctly even after it was changed or interfaced with
other software.
These two techniques are suitable for fixing bugs that are detected right after they are introduced. 
However, if a bug that lied in a codebase for some time is detected, e.g. because of extending the coverage of the tests, a technique like RPA need to be used. That is, RPA and regression testing/continous integration testing are mutually orthogonal techniques

SZZ~\cite{szz,szz2} is an algorithm for identifying commits in a VCS that introduced bugs, however it works in a quite different settings. It assumes, that the bug has been already fixed and that the commit that fixed the bug is explicitly known or can be found in a log file. This allows to identify particular lines of code that fixed the bug and this information is then exploited while searching for the bug-introducing commit. In our settings, the bugs are not fixed yet, thus SZZ cannot be used.  

Finally, we relate the regression predecessors problem with well known problems from graph theory. The latest regression point can be found using the breadth-first-search (BFS) algorithm\cite{graphs}. As our goal is to minimize the number of validity queries, BFS is not suitable as it queries every vertex. Therefore, we come with a new, specialized, algorithm.

\begin{figure*}[!t]
\centering
\begin{minipage}[t]{.48\textwidth}
  \centering
  \includegraphics[scale=0.52]{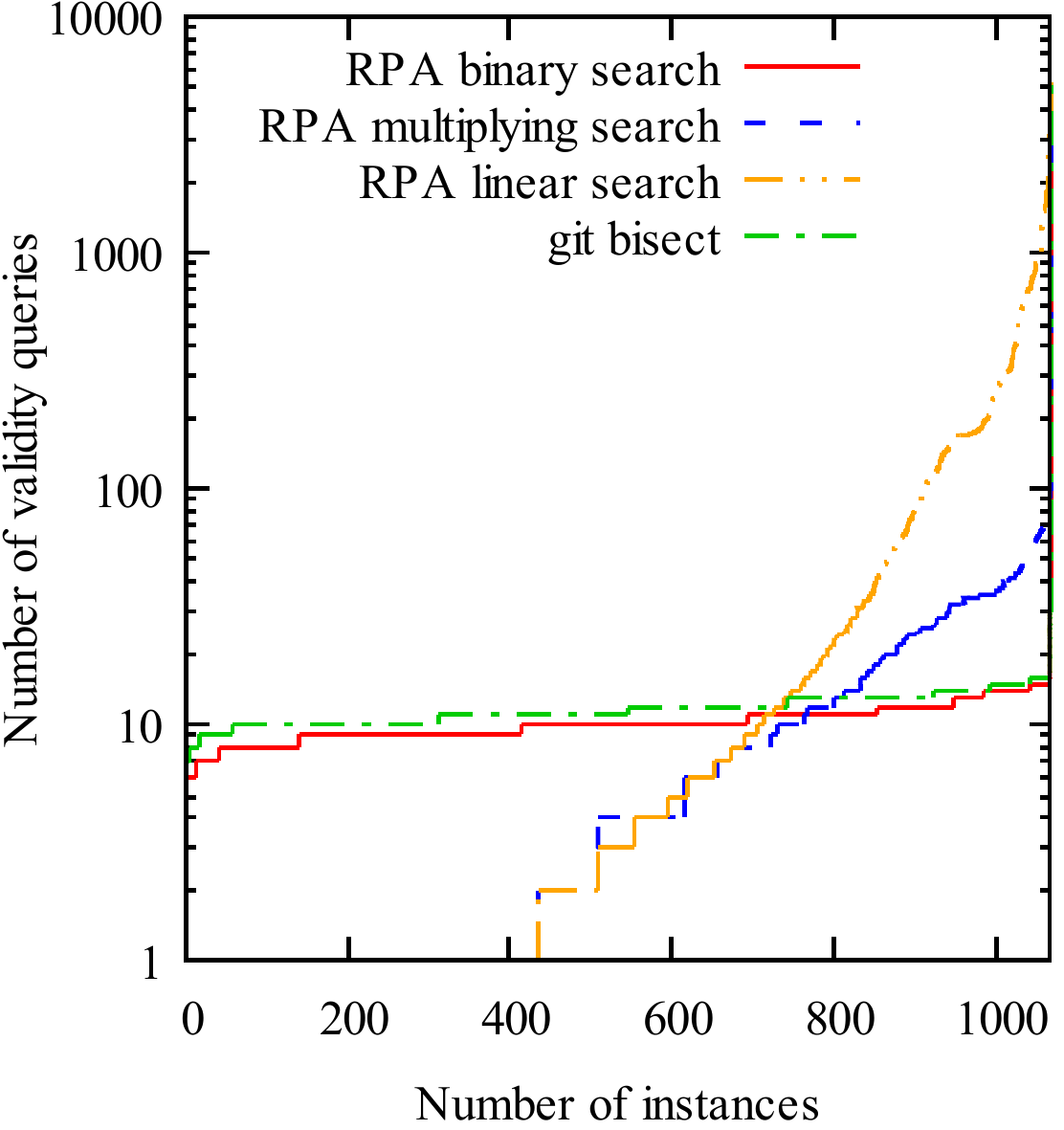}
\caption{Cumulative distribution plot of performed validity queries for each evaluated algorithm. A point with coordinates [x,y] can be read as ``\emph{x} instances were solved by using at most \emph{y} validity queries''.}
\label{res:checks}
\end{minipage}\hspace{10pt}
\begin{minipage}[t]{.48\textwidth}
  \centering
  \includegraphics[scale=0.52]{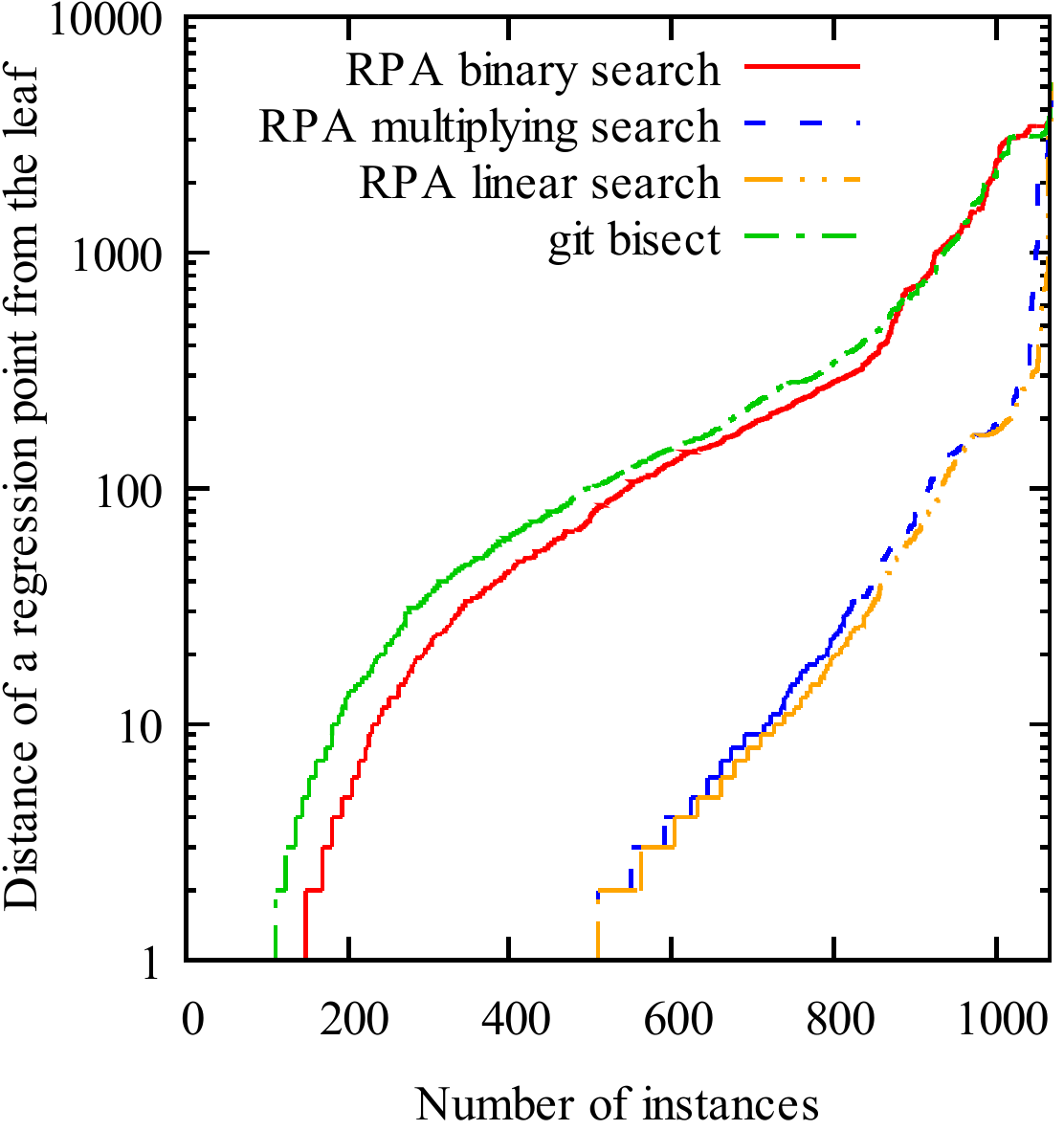}
\caption{Cumulative distribution plot of distance between regression predecessor and corresponding invalid leaf for each evaluated algorithm. A point with coordinates [x,y] can be read as ``in \emph{x} instances the distance was at most \emph{y}''.}
\label{res:dist}
\end{minipage}
\end{figure*}

\section{Experimental Results}\label{sec:exp}
We demonstrate the performance of the variants of RPA on two types of use
cases.  We first focus on the problem of finding a regression predecessor of a
single invalid leaf. We then focus on the problem of finding regression
predecessors of a set of invalid leaves. We also compare the performance of the
RPA variants to that of the git bisect
tool~\cite{git_bisect_doc,git_bisect_alg}.

As  benchmarks we  use large real open source projects, taken from  the GitHub open source
showcases~\cite{showcases},   with at least 8 active branches or at least 1000 commits. Due to the size of the projects it would be
intractable to build and test  all commits in these projects. Therefore  we use those projects from~\cite{showcases} that employ TravisCI~\cite{travis}. Travis CI is a service used to build and test projects hosted
at GitHub and the results of all tests that were run on these projects are
publicly available. Whenever our algorithm needs to validate a commit, it acquires the
results of the tests from the publicly available Travis CI database instead. Overall we
selected 84 projects with  1069
invalid leaves in total. Our selection includes for example the Rails web-application
framework~\cite{rails,rails_paper}, the PHP Interpreter~\cite{php}, or the
ArangoDB~\cite{arangodb}.

We do not provide any details about the architecture of the computer on which
we run the experiments because the computation time is not a relevant criterion
in our study. It took a few seconds to run all the experiments because we didn't actually run the tests. As a main
criterion for measuring the efficiency of evaluated algorithms we use the
number of performed validity queries and the distance between identified
regression points and corresponding invalid leaves (in order to measure the
tendency to find the latest regression points). Complete results of all
measurements are available at \url{https://tinyurl.com/y857h82g}.

\subsection{Single Invalid Leaf Instances}
We first analyse how variants of RPA and git bisect perform while searching for
a regression predecessor of a single invalid leaf. In the case of finding a regression predecessor of a single invalid leaf, it
makes no sense to use propagation. Therefore, we always build the
shortest path from a~valid vertex and employ either binary or multiplying search. We also include the naive linear search approach that builds a path and checks one by one individual commits on the path.

The results comparing the number of the performed validity queries are shown in
Fig.~\ref{res:checks}. In this plot, we show the cumulative distributions of
the performed validity queries for each evaluated algorithm.  The performance
of git bisect and binary search was quite stable on all instances, they needed to
perform at most 17 and 16 validity queries, respectively, to solve the hardest
instances.  On the other hand, the performance of multiplying search was less
stable, it needed to perform from 1 to 101 validity queries. Linear search was negligibly better than exponential search on instances where the regression point was quite close, however it needed to perform up to 3458 validity queries on the harder instances.

\begin{table}[!t]
\centering

\begin{minipage}{.48\textwidth}
\centering
\begin{tabular}{ c | r | r | r}
  $<$ 	& \multicolumn{1}{|c|}{mult}	& \multicolumn{1}{|c|}{bin}	& \multicolumn{1}{|c}{git}\\ \hline
  mult 	& \multicolumn{1}{|c|}{---}	& 736 (69\%)	& 759 (71\%)	\\ \hline
  bin 	& 313 (29\%)			& \multicolumn{1}{|c|}{---}	& 824 (77\%)	\\ \hline
  git	& 294 (28\%)			& 25 (2\%)			& \multicolumn{1}{|c}{---}				\\
\label{res:checks_table}
\end{tabular}
\vspace{10pt}
\caption{The number of instances on which the algorithm named in the row performed strictly less validity queries than the other algorithms.}
\end{minipage}\hfill
\begin{minipage}{.48\textwidth}
\centering
\begin{tabular}{ c | r | r | r}
  $<$ 	& \multicolumn{1}{|c|}{mult}	& \multicolumn{1}{|c|}{bin}	& \multicolumn{1}{|c}{git}\\ \hline
  mult 	& \multicolumn{1}{|c|}{---}	& 717 (67\%)	& 843 (79\%)	\\ \hline
  bin 	& 26 (2\%)			& \multicolumn{1}{|c|}{---}	& 511 (48\%)	\\ \hline
  git	& 27 (3\%)			& 328 (30\%)			& \multicolumn{1}{|c}{---}				\\
\label{res:dist_table}
\end{tabular}
\vspace{10pt}
\caption{The number of instances on which the algorithm named in the row found a~strictly closer regression predecessor than the other algorithms.}
\end{minipage}\hfill
\end{table}

In addition to the plot of cumulative distribution we show in Table~1
the number of instances on which one algorithm performed strictly less validity queries than its competitors. 
The table shows that binary search was superior to git
bisect on most of the instances. This is caused by the nature of these two
algorithms. Both of them need to perform just logarithmically many validity
queries, but git bisect searches the whole graph whereas binary search
traverses only a single path.
Moreover, RPA always chooses
the shortest usable path. 
Multiplying search can execute less than logarithmically many validity queries but it can also execute more than 
logarithmically many validity queries. Its performance depends on the position
of regression points on the path. The closer the regression points are to the
leaf the more likely the multiplying search approach outperforms the binary search approach (and thus
also git bisect).

Besides the number of performed validity queries, we also measure the tendency
of the algorithms to find the latest regression predecessors.
So far we have not precisely defined which regression predecessor is the latest
one and there is more than one suitable definition. In the case of binary,
 multiplying, and linear search we look for a regression predecessor on a path; thus, we can
say that the latest regression predecessor is the regression point which is
closest to the end of the path (i.e., closest to the leaf). In our experiments,
multiplying search found the closest regression predecessor on the path in 86
percent of instances whereas binary search only in 29 percent of instances.
This notion of latest regression predecessor is not applicable to git bisect
because git bisect does not operate on paths. In order to compare RPA with git
bisect, we measured the distance between the found regression predecessor and
the corresponding invalid leaf, i.e.~the shortest path between these two
vertices in the commit graph. Fig.~\ref{res:dist} shows a plot of cumulative
distributions of distance between regression predecessors and corresponding
leaves for each evaluated algorithm. 
The results demonstrate that multiplying
search substantially outperforms binary search and git bisect, and that binary
search is slightly better than git bisect. The naive linear search is just negligibly better than multiplying search.

In addition, Table~2 shows the number of instances on which one algorithm found strictly closer regression point than its competitors. 
The multiplying search strictly dominates
both its competitors; the regression predecessor found by multiplying search
was closer to the leaf than the one found by git bisect in 79 percent of
instances. Binary search performed slightly better, it was dominated by
multiplying search only in 67 percent of instances. 

\begin{figure*}[!t]
\centering
\begin{minipage}[t]{.48\textwidth}
  \centering
\includegraphics[scale=0.52]{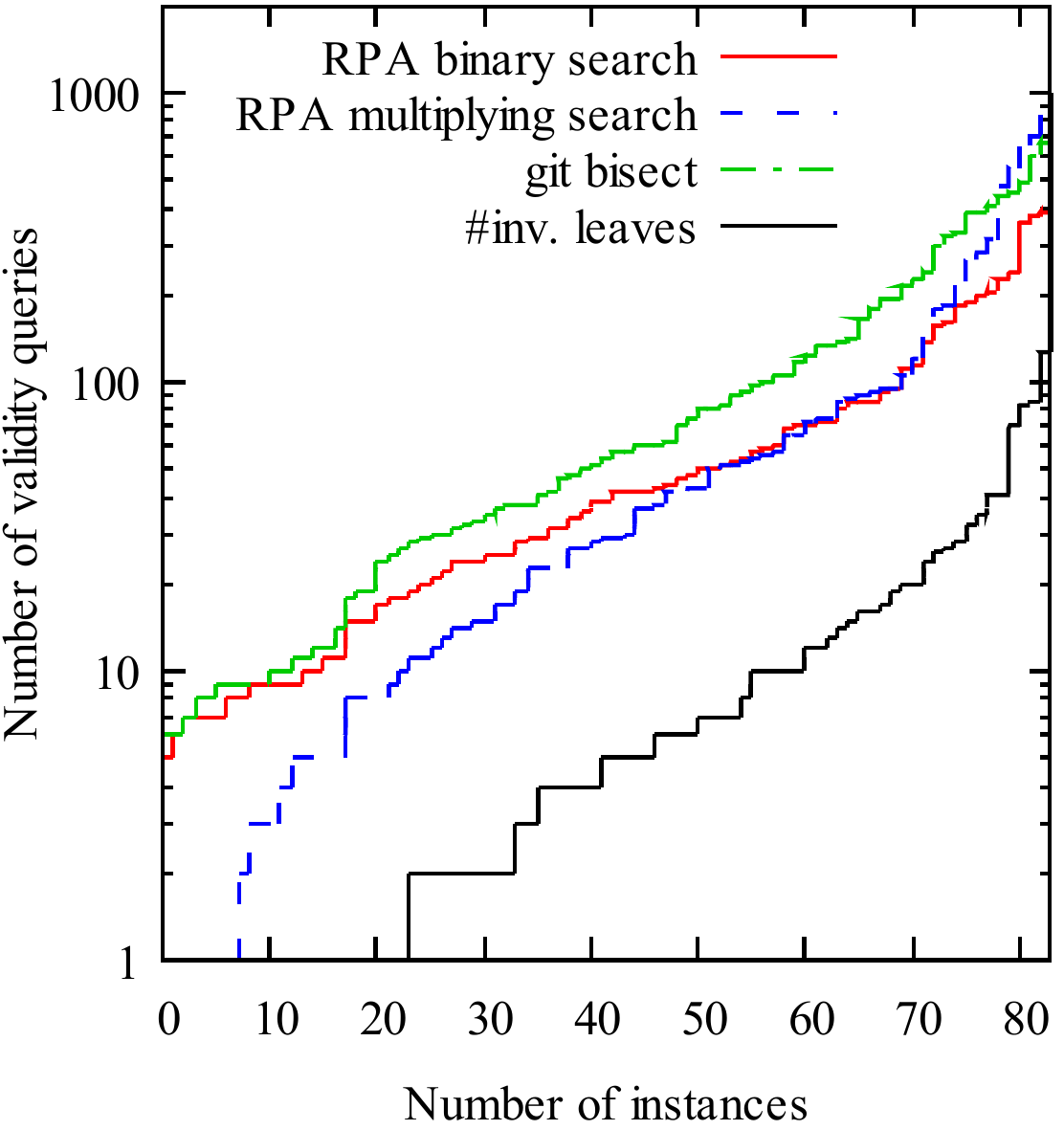}
\caption{Cumulative distributions of number of performed validity queries for git bisect and variants of RPA without propagation. The black line is the cumulative distribution of number of invalid leaves.} \label{res:dist_set_no_prop}
\end{minipage}\hspace{10pt}
\begin{minipage}[t]{.48\textwidth}
  \centering
\includegraphics[scale=0.52]{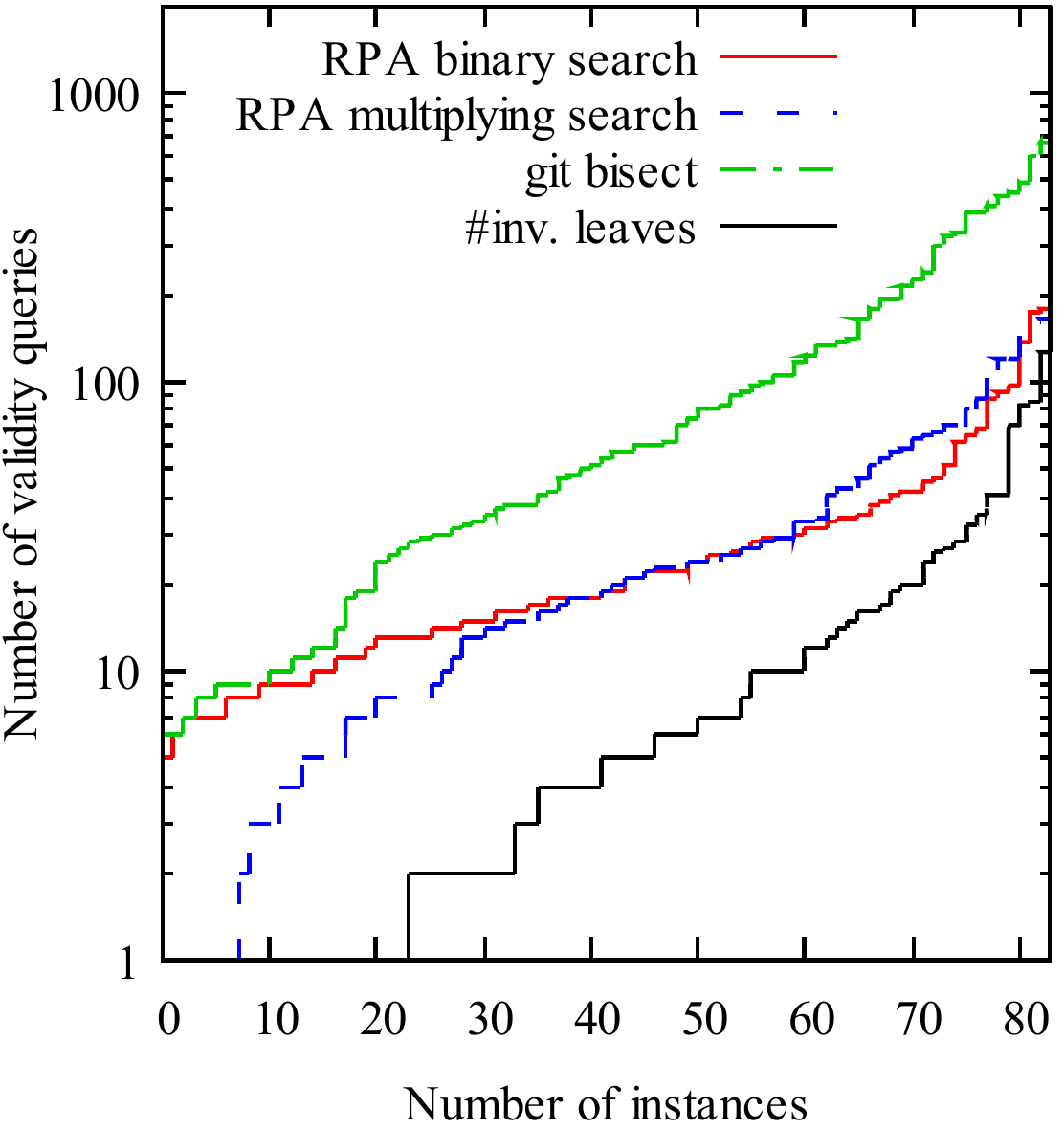}
\caption{Cumulative distributions of number of performed validity queries for git bisect and variants of RPA with propagation. The black line is the cumulative distribution of number of invalid leaves.} \label{res:dist_set_prop}
\end{minipage}
\end{figure*}
\subsection{Sets of Invalid Leaves }
We now demonstrate the performance of the RPA variants on the problem of
finding regression predecessors for a set of invalid leaves.
In particular, we evaluate both proposed approaches for finding regression points, i.e., the binary and multiplying search. Moreover, we evaluate both these approaches in two variants: with and without regression point propagation (the optional part of RPA). 

We also compare the variants of RPA to the git bisect tool. As mentioned in
Sec.~\ref{sec:related}, git bisect deals with the problem of finding a
regression predecessor of a single invalid leaf.  Therefore, in order to solve
the problem of regression predecessors for a given set of invalid leaves $L$,
git bisect has to be run once per each leaf from $L$. As all these runs are
independent, it might happen that some commits are evaluated repeatedly. In
order to avoid the repeated evaluations, we supplement git bisect with a cache
saving the results of the previous evaluations.  Thus, every commit is
evaluated at most once.

As benchmarks we used the 84 projects from GitHub showcases; the goal was to
find a regression predecessor for every invalid leaf in every project.
In this part of experimental evaluation we focus solely on the number of
performed validity queries. Figures~\ref{res:dist_set_no_prop} and
\ref{res:dist_set_prop} show the cumulative distribution plots of the performed
validity queries for the variants of RPA with and without propagation,
respectively. In both plots we also include the results achieved by git bisect.
Moreover, we include the cumulative distribution of the number of invalid
leaves (solid red line), i.e.~a point with coordinates $[x,y]$ means that $x$
instances have at most $y$ invalid leaves.

In general, the regression point propagation significantly reduces the overall number
of performed validity queries.
Considering the difference in performance between variants of RPA with
multiplying and binary search, respectively, we observe the same behaviour as
in the case of finding regression predecessors for single invalid leaves.
There are some instances on which multiplying search outperformed binary search,
and some instances on which binary search outperformed multiplying search. Git
bisect performed conclusively the worst of all evaluated algorithms.

\subsection{Recommendations}
We have presented several variants of the RPA algorithm and the experimental results show that the variants are in general
incomparable. There is no variant that would beat all the others independently of
the comparison criteria. 
In the case where user searches for a regression predecessor of a single invalid leaf it makes no sense to use propagation. We suggest the user to use multiplying search if she prefers finding the closest regression points, and to use binary search if she prefers minimizing the number of performed validity queries. 

In the other case, where user searches for regression predecessors of several invalid leaves, it might be worth to use propagation. If the user prefers finding the closest regression points to minimizing the number of performed validity queries, we suggest her not to use regression point propagation and employ multiplying search. 
In the opposite case, when user focus mainly on minimizing the number of validity queries, we suggest her to use regression point propagation and employ binary search as it guarantees that only logarithmically many validity queries will be performed.

\section{Conclusion}\label{sec:concl}
We present a new algorithm, called the Regression Predecessors Algorithm (RPA),
for finding regression points in projects under version control. The algorithm has
several variants, the choice of which depends on whether we prefer to minimise the
number of validity queries or the latest regression point. We have experimentally
compared the variants among themselves as well as against the state-of-the
art tool git bisect. The results show that the variants of RPA are in general
incomparable as there is no variant that would beat all others independently of
the criteria. The main strength of RPA lies in the ability to minimise the number
of validity queries while respecting the requirement to find the latest regression
point. In all cases RPA is superior to the algorithm used in git bisect.

\bibliographystyle{plain}
\bibliography{refs}

\end{document}